\documentclass[preprint,amsmath,amssymb,aps,superscriptaddress]{revtex4-1}
\usepackage[latin9]{inputenc}
\usepackage{bm}
\usepackage{dcolumn}
\usepackage{color}
\usepackage{hyperref}
\hypersetup{
    bookmarks=true,         
    unicode=false,          
    pdftoolbar=true,        
    pdfmenubar=true,        
    pdffitwindow=false,     
    pdfstartview={FitH},    
    pdftitle={My title},    
    pdfauthor={Author},     
    pdfsubject={Subject},   
    pdfcreator={Creator},   
    pdfproducer={Producer}, 
    pdfkeywords={keyword1} {key2} {key3}, 
    pdfnewwindow=true,      
    colorlinks=true,        
    linkcolor=blue,      
    citecolor=blue,         
    filecolor=blue,      
    urlcolor=blue           
}

\newcommand{\unit}[1]{\ensuremath{\bm{\widehat{\mathrm{#1}}}}}

\newcommand{\epsb}{\epsilon_B}
\newcommand{\fr}{\nu_{r}}
\newcommand{\fk}{\nu_{k}}

\newcommand{\eero}[1]{{#1}}

\makeatother

\begin{document}

\title{\eero{Guiding-center transformation of the radiation-reaction force in a nonuniform magnetic field}}

\author{E. Hirvijoki}
\email{eero.hirvijoki@chalmers.se}
\affiliation{Department of Applied Physics, Chalmers University of Technology, 41296 Gothenburg, Sweden}

\author{J. Decker}
\affiliation{Ecole Polytechnique F\'ed\'erale de Lausanne (EPFL), \\Centre de Recherches en Physique des Plasmas (CRPP), CH-1015 Lausanne, Switzerland}

\author{A. J. Brizard}
\affiliation{Department of Physics, Saint Michael's College, Colchester, VT 05439, USA}

\author{O. Embreus}
\affiliation{Department of Applied Physics, Chalmers University of Technology, 41296 Gothenburg, Sweden}

\date{December 5, 2014}

\begin{abstract}
In this paper, we present the guiding-center transformation of the \eero{radiation-reaction force of a classical point charge} traveling in a nonuniform magnetic field. The transformation is valid as long as the gyroradius of the charged particles is much smaller than the magnetic field nonuniformity length scale, so that the guiding-center Lie-transform method is applicable. 
Elimination of the gyromotion time scale from the \eero{radiation-reaction} force is obtained with the Poisson bracket formalism originally introduced by {[}A. J. Brizard, Phys. Plasmas \textbf{11} 4429 (2004){]}, where it  was used to eliminate the fast gyromotion from the Fokker-Planck collision operator. The formalism presented here is applicable to the motion of charged particles in planetary magnetic fields as well as in magnetic confinement fusion plasmas, where the corresponding so-called synchrotron radiation can be detected. Applications of the guiding-center radiation-reaction force include tracing of charged particle orbits in complex magnetic fields as well as kinetic description of plasma when the loss of energy and momentum due to radiation plays an important role, e.g., for runaway electron dynamics in tokamaks.
\end{abstract}

\pacs{put Pacs here}

\maketitle

\section{Introduction}
\label{sec:intro} An accelerated charged particle emits electromagnetic radiation and loses energy and momentum in reaction, in accordance with the \eero{radiation-reaction force (RR-force)~\cite{lorentz,abraham,dirac:1938,pauli1958theory,landau_lifshitz_fields}}. In a magnetized plasma, the radiation resulting from the gyration around the field lines is often referred to as the synchrotron emission. The RR-force increases with the particle energy and the accelerating force, which itself depends on the velocity in the case of the Lorentz force. The effect of the radiation on the particle motion can be significant for very energetic particles. In particular, it contributes to limit the energy reached by runaway electrons in tokamak plasmas~\cite{Bakhtiari_et_al:PRL2005}, \eero{and explains the observation of an elevated critical electric field for runaway-electron generation~\cite{adam:effective_field_PRL}}.

One of the consequences of magnetic field nonuniformity in axisymmetric configurations such as dipole or tokamak fields is the superposition of three periodic motions in the particle trajectories, namely the gyromotion, bounce or transit motion, and drift precession. The acceleration associated with each periodic motion would in turn contribute to the radiation losses. Another consequence of the magnetic nonuniformity is that the radiation \eero{RR-force} could induce transport of particles across the magnetic flux-surfaces. As the \eero{RR-force} increases with increasing particle energy while the collisional force decreases with velocity, radial transport associated with the \eero{RR-force} could overcome collisional transport at high relativistic energies.

In the presence of a weak magnetic nonuniformity such that the gyroradius of the charged particle is much smaller than the magnetic field nonuniformity length scale, the magnetic moment, $\mu$, is an adiabatic invariant of the Hamiltonian particle motion. In this case, it is often useful to separate the gyromotion from the rest of the particle motion, and study longer time scales. Making use of the adiabatic invariance, Lie-transform perturbation methods can be used to eliminate the fast gyromotion and to derive the underlying guiding-center dynamics. Cumulative effects of the gyromotion in a nonuniform magnetic field, such as the mirror force and magnetic drifts, are entirely retained in the guiding-center dynamics, which is typically much easier to compute than the full particle dynamics. This approach is one of the classical results in modern plasma physics~\cite{littlejohn:jpp:4732464} and has been summarized in a review paper~\cite{RevModPhys.81.693}. 

Previous attempts to include the effect of magnetic field nonuniformity in the \eero{RR-force} were made without going through a proper guiding-center transformation. In a first paper (Ref.~\cite{andersson:pop2001}), a contribution from the magnetic field-line curvature was argued in addition to the uniform-field formulation. This corrective term, which is of second order in magnetic-field nonuniformity, is interesting since it does not vanish for $\mu\rightarrow 0$. Their approach, however, neglects a number of terms that are of first order in magnetic-field nonuniformity. Furthermore, some of the first order terms and all second order terms in the Hamiltonian motion are neglected which leads to inconsistent treatment of the dissipative and Hamiltonian guiding-center dynamics. \eero{One can understand how keeping only the leading-order terms is necessary for conducting analytical studies of complex phemomena, as done in Ref.~\cite{andersson:pop2001}, but in particle following applications where the trajectory is calculated numerically one should be as consistent as possible to prevent, e.g., numerical drift of the energy due to improper equations.} In Refs.~\cite{guan:2010,liu:pop2014}, a rather different approach is adopted by treating the \eero{RR-force} as an ''effective electric field'' which is then added into guiding-center Lagrangian as a time depending perturbation in the vector potential. This is not correct: the \eero{RR-force} is of dissipative nature and no practical Lagrangian formulation exists within the framework of classical electrodynamics (here we do not discuss the quantum mechanical treatments). Moreover, the ''effective field'' in Refs.~\cite{guan:2010,liu:pop2014} is given only for a simple toroidal geometry and if one calculates the corresponding equations of motion using the Euler-Lagrange equation, the result does not give the ''effective electric field'' that they start with. 

As the radiatiative momentum losses, however, are important for the dynamics of relativistic charged particles, we see that a guiding-center describtion that is consistent with the Hamiltonian formalism is necessary. In the present paper, we derive the guiding-center \eero{RR-force} in a weakly nonuniform magnetic field using Lie-transform perturbation methods. 
In Sec.~\ref{sec:particle_force}, we first introduce the particle phase-space \eero{RR-force} and give its expression in a nonuniform magnetic field. A general method for including non-Hamiltonian forces into guiding-center formalism is described in Sec.~\ref{sec:dissipation}. The guiding-center transformation is carried out explicitly in Sec.~\ref{sec:gc_radiation_reaction}, where corrections to the guiding-center equations of motion arising from the radiation losses are derived consistently with the first order guiding-center theory. Applications of the guiding-center \eero{RR-force} are discussed in the conclusion.

\newpage
\section{Radiation-Reaction force in a nonuniform magnetic field}
\label{sec:particle_force}
\eero{
The radiation-reaction force was first described for a classical non-relativistic point charge by Lorentz~\cite{lorentz}. Later, Abraham~\cite{abraham} and Dirac~\cite{dirac:1938} generalized it to relativistic energies obtaining the Lorentz--Abraham--Dirac (LAD) force~\cite{pauli1958theory}
\begin{align}
\mathbf{K}=\frac{e^{2}\gamma^{2}}{6\pi\varepsilon_{0}c^{3}}\left[\ddot{\mathbf{v}}+\frac{3\gamma^{2}}{c^{2}}\left(\mathbf{v}\cdot\dot{\mathbf{v}}\right)\dot{\mathbf{v}}+\frac{\gamma^{2}}{c^{2}}\left(\mathbf{v}\cdot\ddot{\mathbf{v}}+\frac{3\gamma^{2}}{c^{2}}\left(\mathbf{v}\cdot\dot{\mathbf{v}}\right)^{2}\right)\mathbf{v}\right],\label{eq:reaction_force}
\end{align}
where $e$ is the particle charge, $\gamma=1/\sqrt{1-v^{2}/c^{2}}=\sqrt{1+p^{2}/(mc)^{2}}$
is the relativistic factor and $\mathbf{p}=\gamma m\mathbf{v}$ is
the particle momentum. The LAD-force, however, contains third order time derivatives with respect to particle position and cannot be uniquely solved given initial values for the particle position and velocity, therefore violating causality. Another well-known problem with the LAD-force is that in the absence of external forces it allows the existence of so-called {\em runaway solutions} (see, e.g., Ref.~\cite{rohrlich2007classical}) that lead to exponential growth of the particle velocity. These issues have generated discussion regarding what expression to use for the RR-force (see, e.g., the excellent text in Ref.~\cite{griffiths:AJP:2009}). Landau and Lifshitz~\cite{landau_lifshitz_fields}, for example, suggest a perturbative approach where the velocity derivatives in Eq.~(\ref{eq:reaction_force}) are expressed in terms of the external force. Ford and O'Connel~\cite{Ford1993182} say that this approach is in fact the correct one. In the paper by Spohn~\cite{Spohn:EPL:2000}, it is shown that LAD-force should be limited on a so-called {\em critical surface} to avoid unphysical solutions, and that this actually corresponds to the perturbative approach. In this paper, we have chosen to use the Landau-Lifshitz formula, as it does not suffer from the unphysical behaviour associated with the LAD-force.

In magnetized plasmas, the particle motion is typically dominated by the magnetic force $\mathbf{F}_{m}=e\,\mathbf{v}\times\mathbf{B}$. If we neglect the electric field, and replace the velocity derivatives in Eq.~(\ref{eq:reaction_force}) with the Lorentz force, the expression for the RR-force simplifies to~\cite{landau_lifshitz_fields}
\begin{align}
\mathbf{K}\;= & \;-\fr\left(\mathbf{p}_{\bot}+\frac{p_{\bot}^{2}}{(mc)^{2}}\mathbf{p}\right)-\epsilon\,\fr\Omega^{-1}B^{-1}\dot{\mathbf{B}}\times\mathbf{p},\label{eq:Kapprox}
\end{align}
where the Larmor frequency is $\Omega=eB/(\gamma m)$. The perpendicular momentum in Eq.~(\ref{eq:Kapprox}) is $\mathbf{p}_{\bot}=(\mathbf{I}-\unit{b}\unit{b})\cdot\mathbf{p}$, with the magnetic field unit vector being $\unit{b}=\mathbf{B}/B$. The characteristic time for RR-force is  
\begin{align}
\fr^{-1}=\frac{6\pi\varepsilon_{0}\gamma(mc)^{3}}{e^{4}B^{2}}=\frac{3c}{2\gamma r_e\Omega^2},
\end{align}
where $r_e=e^2/(4\pi\varepsilon_0mc^2)$ is the classical electron radius. With this notation $\fr/\Omega=(2r_e\gamma\Omega)/(3c)$, which is typically much smaller than one (for electrons $\fr/\Omega\sim 10^{-12} B / [T]$). It is thus obvious that the magnetic force $|\mathbf{F}_m|\sim\Omega p$ dominates the RR-force $|\mathbf{K}|\sim\fr p$, and that the RR-force therefore acts a dissipation to the Hamiltonian motion in the external magnetic field.

In equation~(\ref{eq:Kapprox}), we have also introduced the dimensionless parameter $\epsilon$ as the dimensionless guiding-center ordering parameter (renormalized charge $e\rightarrow \epsilon^{-1}e$) which will be used throughout the guiding-center transformation. Physical results are obtained by setting~$\epsilon=1$.
}
\section{Dissipative forces in Guiding-center formalism}
\label{sec:dissipation} To transform general dissipative forces into guiding-center
formalism, we cannot apply the Lie-transformation exactly as it is done for deriving the Hamiltonian guiding-center equations of motion: General dissipative forces do not necessarily have a phase-space Lagrangian formulation. Instead, we proceed via detour. Transforming the particle phase-space
continuity equation we will identify the components of the guiding-center
force for any desired combination of phase space coordinates and simultaneously
guarantee the density conservation. This approach can be further used to include the \eero{RR-force} into the guiding-center kinetic equation.

Starting from the particle phase-space continuity equation 
\begin{equation}
\frac{\partial f}{\partial t}+\frac{\partial}{\partial\mathbf{z}}\cdot\left(\frac{}{}\dot{\mathbf{z}}f\right)+\frac{\partial}{\partial\mathbf{p}}\cdot\left(\frac{}{}\mathbf{K}f\right)=0,
\end{equation}
where $\mathbf{z}=(\mathbf{x},\mathbf{p})$ \eero{are the phase-space coordinates} and $\mathbf{K}$ now
denotes a general dissipative force, we first express the continuity
equation in terms of the charged particle Hamiltonian 
\begin{equation}
H=\gamma mc^{2},
\end{equation}
and the non-canonical charged particle Poisson bracket 
\begin{equation}
\label{eq:bracket_definition}
\{f,g\}=\frac{\partial f}{\partial\mathbf{x}}\cdot\frac{\partial g}{\partial\mathbf{p}}-\frac{\partial f}{\partial\mathbf{p}}\cdot\frac{\partial g}{\partial\mathbf{x}}+e\mathbf{B}\cdot\frac{\partial f}{\partial\mathbf{p}}\times\frac{\partial g}{\partial\mathbf{p}},
\end{equation}
to obtain the Poisson-bracket formulation of the particle phase-space
continuity equation 
\begin{equation}
\frac{\partial f}{\partial t}+\{f,H\}+\{x^{i},K_{i}\, f\}=0.
\end{equation}
Here $x^{i}$ is the Cartesian component of the position $\mathbf{x}$,
and summation over repeated indices is assumed. We have also made use of the identity
\begin{align}
\dot{z}^{\alpha}\equiv\{z^{\alpha},H\},
\end{align}
\eero{that describes the Hamiltonian contribution to the equations of motion}, and of the properties of the Poisson bracket
\begin{align}
\{f,g\}&\equiv\frac{\partial f}{\partial z^{\alpha}}\{z^{\alpha},z^{\beta}\}\frac{\partial g}{\partial z^{\beta}}\equiv\frac{1}{\mathcal{J}}\frac{\partial}{\partial z^{\alpha}}\left(\frac{}{}\mathcal{J}\{f,z^{\alpha}\}\, g\right),
\end{align}
where $\mathcal{J}$ is the phase-space Jacobian satisfying $d\mathbf{z}=\mathcal{J}d^6z$.

The guiding-center
transformation of the particle phase-space continuity equation is
then given by applying the guiding-center push-forward~$\mathcal{T}_{gc}^{-1}$
according to 
\begin{align}
 & \mathcal{T}_{gc}^{-1}\left(\frac{\partial f}{\partial t}+\{f,H\}+\{x^{i},K_{i}\, f\}\right)=0\nonumber \\
 & \Rightarrow\quad\frac{\partial F}{\partial t}+\left\{F,H_{gc}\right\}_{gc}+\left\{\mathcal{T}_{gc}^{-1}x^{i},(\mathcal{T}_{gc}^{-1}K_{i})\, F\right\}_{gc}=0,
\end{align}
where $F\equiv\mathcal{T}_{gc}^{-1}f$ and $\{\,\cdot\,,\,\cdot\,\}_{gc}\equiv\mathcal{T}_{gc}^{-1}\{\mathcal{T}_{gc}\,\cdot,\mathcal{T}_{gc}\,\cdot\}$ are now the guiding-center
distribution function and Poisson-bracket, respectively.
For an explanation of the transformation rules, we encourage the reader
to Ref.~\cite{brizard:4429}. Expressing the guiding-center Poisson
bracket in a divergence form we obtain 
\begin{equation}
\frac{\partial F}{\partial t}+\frac{1}{\mathcal{J}_{gc}}\frac{\partial}{\partial Z^{\alpha}}\left[\mathcal{J}_{gc}\left(\frac{}{}\dot{Z}^{\alpha}+\left\{\mathcal{T}_{gc}^{-1}x^{i},Z^{\alpha}\right\}_{gc}(\mathcal{T}_{gc}^{-1}K_{i})\right)F\right]=0,
\end{equation}
where $Z^{\alpha}$ are the guiding-center phase-space coordinates
and $\mathcal{J}_{gc}$ is the guiding-center phase-space Jacobian, $d\mathbf{Z}=\mathcal{J}_{gc}d^{6}Z$.

If the characteristic time-scale, $\fk^{-1}$, of the dissipative
force, $|\mathbf{K}|\sim\fk p$, is much longer than the
time-scale related to the gyromotion, i.e., $\fk/\Omega\ll1$,
a closure scheme to obtain an equation for a gyroaveraged distribution
function, $\left\langle F\right\rangle $, proceeds in a similar
manner as was presented for the collision operator in Ref.~\cite{brizard:4429}.
To lowest order in $\epsilon_{\nu}=\fk/\Omega$, the gyro-averaged
guiding-center continuity equation thus becomes 
\begin{equation}
\frac{\partial\left\langle F\right\rangle }{\partial t}+\frac{1}{\mathcal{J}_{gc}}\frac{\partial}{\partial Z^{\alpha}}\left[\mathcal{J}_{gc}\left(\frac{}{}\dot{Z}^{\alpha}+\left\langle\left\{\mathcal{T}_{gc}^{-1}x^{i},Z^{\alpha}\right\}_{gc}(\mathcal{T}_{gc}^{-1}K_{i})\right\rangle\right)\left\langle F\right\rangle \right]=0,
\end{equation}
where we have used the fact that the Hamiltonian guiding-center equations
of motion $\dot{Z}^{\alpha}$ are, by construction, independent of
the gyro-angle.

From the gyro-averaged guiding-center continuity equation, we finally
obtain an expression for a dissipative gyro-averaged guiding-center
force
\begin{equation}
\mathcal{K}_{gc}^{\alpha}=\left\langle \frac{}{}\{\mathcal{T}_{gc}^{-1}x^{i},Z^{\alpha}\}_{gc}(\mathcal{T}_{gc}^{-1}K_{i})\right\rangle \equiv\left\langle \Delta^{i\alpha}\cdot\mathcal{T}_{gc}^{-1}K_{i}\right\rangle ,
\end{equation}
where $\left\langle \,\dots\,\right\rangle =1/(2\pi)\int_{0}^{2\pi}d\theta\,\dots$
is the gyro-angle average, and 
\begin{equation}
\bm{\Delta}^{\alpha}\equiv\unit{e}_{i}\Delta^{i\alpha}\equiv\unit{e}_{i}\{\mathcal{T}_{gc}^{-1}x^{i},Z^{\alpha}\}_{gc},
\end{equation}
is the so-called guiding-center projection coefficient.

\section{First order transformation in $(\mathbf{X},p_{\parallel},\mu)$ phase-space}
\label{sec:gc_radiation_reaction} As the time scale related to the radiation reaction satisfies the condition $\fr/\Omega\ll1$,
we can calculate the guiding-center transformation of the \eero{RR-force} according to the method described in Sec.~\ref{sec:dissipation}. In order to proceed, explicit expressions for the projection coefficients~$\Delta^{i\alpha}$ are derived in Appendix~\ref{sec:gc_analysis}.

Considering the phase-space $(\mathbf{X},p_{\parallel},\mu)$, often used in particle tracing, we obtain 
\begin{align}
\Delta^{ij}\;= & \;-\epsilon\frac{\unit{b}}{eB_{\parallel}^{\star}}\times\left(\unit{e}^{i}+\nabla^{\star}\rho_{\epsilon}^i\right)\cdot\,\unit{e}^{j}-\frac{B^{\star j}}{B_{\parallel}^{\star}}\frac{\partial\rho_{\epsilon}^i}{\partial p_{\parallel}},\\
\Delta^{ip_{\parallel}}\;= & \;\frac{\mathbf{B^{\star}}}{B_{\parallel}^{\star}}\cdot\left(\unit{e}^{i}+\nabla^{\star}\rho^i_{\epsilon}\right),\\
\Delta^{i\mu}\;= & \;\epsilon^{-1}\,\frac{e}{m}\frac{\partial\rho^i_{\epsilon}}{\partial\theta},
\end{align}
where $\epsilon$ is the dimensionless guiding-center ordering parameter and the transformed gyroradius vector is defined as
\begin{align}
\bm{\rho}_{\epsilon}\equiv\mathcal{T}_{gc}^{-1}\mathbf{x}-\mathbf{X}\equiv\epsilon\bm{\rho}_0+\epsilon^2\bm{\rho}_1+\dots,
\end{align}
with the sub-indices referring to the order with respect to magnetic field nonuniformity. Thus, keeping the gyroradius up to the term $\bm{\rho}_1$ gives projection coefficients that are valid up to first order in the magnetic field nonuniformity.

The expressions for the so-called symplectic or effective magnetic field $\mathbf{B}^{\star}$ and the modified gradient operator $\nabla^{\star}$ as well as for the zeroth and first order gyroradius vectors, $\bm{\rho}_0$ and $\bm{\rho}_1$, are given in the Appendix~\ref{sec:gc_analysis}. 

Similarly as for $\bm{\rho}_{\epsilon}$, we have for the push-forward of the particle phase-space \eero{RR-force}
\begin{align}
\mathcal{T}_{gc}^{-1}\mathbf{K}\equiv\mathbf{K}_{\epsilon}=\mathbf{K}_0+\epsilon\mathbf{K}_1+\dots,
\end{align}
where the sub-indices again refer to the order with respect to the magnetic field nonuniformity. Explicit expression for $\mathbf{K}_{\epsilon}$ is given in Appendix~\ref{sec:push_forwards}. Using the expressions for the projection coefficients we then find the components for the guiding-center radiation reaction force
\begin{align}
\mathcal{K}^{\mathbf{X}}\;= & \;-\epsilon\frac{\unit{b}}{eB_{\parallel}^{\star}}\times\left\langle\mathbf{K}_{\epsilon}+\nabla^{\star}\bm{\rho}_{\epsilon}\cdot\mathbf{K}_{\epsilon}\right\rangle -\frac{\mathbf{B}^{\star}}{B_{\parallel}^{\star}}\left\langle\frac{\partial\bm{\rho}_{\epsilon}}{\partial p_{\parallel}}\cdot\mathbf{K}_{\epsilon}\right\rangle,\\
\mathcal{K}^{p_{\parallel}}\;= & \;\frac{\mathbf{B}^{\star}}{B_{\parallel}^{\star}}\cdot\left\langle\mathbf{K}_{\epsilon}+\nabla^{\star}\bm{\rho}_{\epsilon}\cdot\mathbf{K}_{\epsilon}\right\rangle,\\
\mathcal{K}^{\mu}\;= & \;\epsilon^{-1}\,\frac{e}{m}\left\langle \frac{\partial\bm{\rho}_{\epsilon}}{\partial\theta}\cdot\mathbf{K}_{\epsilon}\right\rangle.
\end{align}
More explicitly, the expressions valid up to first order in magnetic field nonuniformity are
\begin{align}
\label{eq:k1}\mathcal{K}^{\mathbf{X}}\;= & \;-\epsilon\frac{\unit{b}}{eB_{\parallel}^{\star}}\times\left\langle\mathbf{K}_{0}+\epsilon\mathbf{K}_{1}+\epsilon\nabla^{\star}\bm{\rho}_{0}\cdot\mathbf{K}_{0}\right\rangle-\epsilon^2\unit{b}\left\langle\frac{\partial\bm{\rho}_{1}}{\partial p_{\parallel}}\cdot\mathbf{K}_{0}\right\rangle,\\
\mathcal{K}^{p_{\parallel}}\;= & \;\frac{\mathbf{B}^{\star}}{B_{\parallel}^{\star}}\cdot\Big\langle\mathbf{K}_{0}\Bigr\rangle+\epsilon\unit{b}\cdot\left\langle\mathbf{K}_1+\nabla^{\star}\bm{\rho}_{0}\cdot\mathbf{K}_{0}\right\rangle,\\
\label{eq:k3}\mathcal{K}^{\mu}\;= & \;\frac{e}{m}\left\langle \frac{\partial\bm{\rho}_{0}}{\partial\theta}\cdot\Bigl(\mathbf{K}_{0} +\epsilon\mathbf{K}_{1}\Bigr)\right\rangle +\epsilon\frac{e}{m}\left\langle \frac{\partial\bm{\rho}_{1}}{\partial\theta}\cdot\mathbf{K}_{0}\right\rangle,
\end{align}
where we have noted that $\bm{\rho}_0$ is independent of $p_{\parallel}$ and that $\nabla^{\star}\bm{\rho}_1$ is of second order in magnetic field nonuniformity. 

Gyro-averages of the expressions in Eq. \ref{eq:k1}-\ref{eq:k3} are carried out in Appendix~\ref{sec:gc_force}. The resulting components of the guiding-center \eero{RR-force}, acting as dissipative terms in the equations of motion for the corresponding coordinate, are, for the guiding-center position:
\begin{equation}
\mathcal{K}^{\mathbf{X}}\;=\;-\epsb\,\frac{\fr}{\Omega_{\parallel}^{\star}}\frac{2\mu B}{mc^2}\left(\unit{b}\times\dot{\mathbf{X}}+3v_{\parallel}\,\varrho_{\parallel}\,\bm{\kappa}\right),
\end{equation}
for the parallel momentum:  
\begin{align}
\label{eq:kpar}
\mathcal{K}^{p_{\parallel}}\; = & \; -\fr\,p_{\parallel}\frac{\mu B}{mc^2}\left(2+\epsb\varrho_{\parallel}\tau_B\right)-\epsb\,\fr\frac{p_{\perp}\gamma^2}{2}\varrho_{\perp}\tau_B,
\end{align}
and for the magnetic moment:
\begin{align}
\label{eq:kmu}
\mathcal{K}^{\mu}\; = & \; -\fr\,\mu\left(1+\frac{2\mu B}{mc^2}\right)\left(2+\epsb\varrho_{\parallel}\tau_B\right),
\end{align}
where the parameter $\epsb$ is introduced to explicitly point out the contribution from magnetic field nonuniformity. The parallel and perpendicular gyroradius and the modified gyro-frequency are defined as
\begin{align}
\varrho_{\parallel}\;= & \;\frac{p_{\parallel}}{eB},\qquad\varrho_{\perp}\;= \;\frac{p_{\perp}}{eB},\qquad\Omega_{\parallel}^{\star}=\Omega\,(1+\epsb\varrho_{\parallel}\tau_B).
\end{align}
The first order corrections in Eqs.~(\ref{eq:kpar}) and~(\ref{eq:kmu}) relate to the magnetic field-line twist parameter, $\tau_B=\unit{b}\cdot\nabla\times\unit{b}$, which also appears in Hamiltonian guiding-center motion as the phase-space Jacobian is $\mathcal{J}_{gc}\equiv B_{\parallel}^{\star}\equiv B\,(1+\epsb\varrho_{\parallel}\tau_B)$.

While the phase-space $(\mathbf{X},p_{\parallel},\mu)$ was used to carry out the explicit guiding-center transformation, properties of the Poisson-bracket (\ref{eq:bracket_definition}) provide general rules for transforming between different phase-spaces according to
\begin{align}
\mathcal{K}_{gc}^{\alpha}=\mathcal{K}^{\mathbf{X}}\cdot\nabla\alpha+\mathcal{K}^{p_{\parallel}}\frac{\partial\alpha}{\partial p_{\parallel}}+\mathcal{K}^{\mu}\frac{\partial\alpha}{\partial\mu},
\end{align}

\section{Conclusions}
For the first time, a consistent guiding-center transformation of the radiation-reaction force \eero{for a particle traveling in an external magnetic field} is presented up to the first order in the magnetic-field nonuniformity. As a result, we observe corrections that are proportional to the magnetic field-line twist, which itself plays an important role in the Hamiltonian equations of motion. As the magnetic moment is an exact invariant of Hamiltonian guiding-center motion, the first order correction especially to the dissipative evolution of the magnetic moment could be important: numerical simulations of tokamak first wall power loads from fast particles are very sensitive to the details of the magnetic field nonuniformities and to the guiding-center phase-space trajectories. Due to the presence of the three periodic motions (gyro, bounce, and precession), small deviations in the guiding-center trajectory may cumulate and change the wall power loads. As magnetic perturbations are considered as an option to mitigate the formation of dangerously large runaway electron beams in tokamaks, equations to model the dissipative guiding-center motion must be accurate and consistent with the rest of the tools.

The equations derived in this paper are applicable also to particle dynamics in astrophysical plasmas as well as to any magnetically confined laboratory plasmas as long as the guiding-center formalism itself is valid. The paper also provides a method for transforming general dissipative forces to guiding-center phase-space. The procedure is valid as long as the relative momentum loss over a gyroperiod is sufficiently small.

Carrying the transformation up to second order in magnetic field non-uniformity would provide an additional term in $\mathcal{K}^{p_{\parallel}}$ that does not vanish for $\mu\rightarrow 0$. This component would further contribute to the guiding-center motion along field-lines in a curved magnetic field, and would dominate in the limit $\mu\rightarrow0$. The corresponding second order guiding-center transformation of the radiation reaction force will be presented in a future contribution. 

As the momentum loss due to the \eero{RR-force} is also typically small over a particle bounce or transit time, performing an orbit averaging operation for axisymmetric configurations as prescribed in Ref. \cite{brizard:2009} would yield a reduced orbit-averaged guiding-center \eero{RR-force} operator in a three-dimensional phase-space. Radial transport coefficents including neoclassical effects could then be explicitly derived. Such operator could then be readily implemented in a 3-D orbit-averaged guiding-center kinetic code \cite{decker:2010}.

\begin{acknowledgments}
We would like to thank Dr. Istvan Pusztai, Mr. Adam Stahl, and Prof. T\"{u}nde F\"{u}l\"{o}p for fruitful discussions on improving the manuscript. Work by A.~J.~Brizard  was supported by a US DoE grant under contract No. DE-SC0006721.
\end{acknowledgments}

\appendix
\section{Relativistic guiding-center transformation}
\label{sec:gc_analysis} The relativistic guiding-center Lagrangian
one-form for the guiding-center phase-space coordinates $Z^{\alpha}=(\mathbf{X},p_{\parallel},\mu,\theta)$
is 
\begin{align}
\Gamma_{gc}\equiv\Gamma_{\alpha}\mathrm{d}Z^{\alpha}-H_{gc}\mathrm{d}t= \left(\epsilon^{-1}e\mathbf{A}+p_{\parallel}\unit{b}\right)\cdot\mathrm{d}\mathbf{X}+\epsilon\frac{m\,\mu}{e}\left(\mathrm{d}\theta-\mathbf{R}^{\star}\cdot\mathrm{d}\mathbf{X}\right)
  -\gamma mc^{2}\mathrm{d}t,
\end{align}
where $\gamma=\sqrt{1+p_{\parallel}^2/(mc)^2+2\mu B/(mc^2)}$ and $\epsilon$ is the guiding-center ordering parameter, the modified
gyrogauge field is $\mathbf{R}^{\star}=\mathbf{R}+(\tau_{B}/2)\unit{b}$
with $\mathbf{R}=\nabla\unit{\perp}\cdot\unit{\rho}\equiv\nabla\unit{1}\cdot\unit{2}$
the Littlejohn's gyrogauge vector, and $\tau_{B}=\unit{b}\cdot\nabla\times\unit{b}$
the magnetic field line torsion. The two right-handed orthogonal unit
vector sets, $(\unit{b}(\mathbf{X}),\unit{\perp}(\mathbf{X},\theta),\unit{\rho}(\mathbf{X},\theta))$
and $(\unit{b}(\mathbf{X}),\unit{1}(\mathbf{X}),\unit{2}(\mathbf{X}))$
are 
\begin{align}
\unit{\rho}\;= & \;\cos\theta\,\unit{1}-\sin\theta\,\unit{2},\\
\unit{\perp}\;= & \;-\sin\theta\,\unit{1}-\cos\theta\,\unit{2}.
\end{align}
The guiding-center Poisson bracket calculated from the guiding-center
one-form $\Gamma_{\alpha}\mathrm{d}Z^{\alpha}$ is 
\begin{align}
\{F,G\}_{gc}\;= & \;\epsilon^{-1}\frac{e}{m}\left(\frac{\partial F}{\partial\theta}\frac{\partial G}{\partial\mu}-\frac{\partial F}{\partial\mu}\frac{\partial G}{\partial\theta}\right)\nonumber \\
&  \;+\frac{\mathbf{B}^{\star}}{B_{\parallel}^{\star}}\cdot\left(\nabla^{\star}F\frac{\partial G}{\partial p_{\parallel}}-\frac{\partial F}{\partial p_{\parallel}}\nabla^{\star}G\right) -\epsilon\,\frac{\unit{b}}{eB_{\parallel}^{\star}}\cdot\nabla^{\star}F\times\nabla^{\star}G,
\end{align}
where the modified gradient operator is $\nabla^{\star}=\nabla+\mathbf{R}^{\star}\partial/\partial\theta$,
the phase-space Jacobian is $\mathcal{J}_{gc}=\mathbf{B}^{\star}\cdot\unit{b}\equiv B_{\parallel}^{\star}$,
and the effective magnetic field is 
\begin{equation}
\mathbf{B}^{\star}=\nabla\times\left(\mathbf{A}+\epsilon\frac{p_{\parallel}}{e}\unit{b}\right)+\mathcal{O}(\epsilon^{2})\;=\;\mathbf{B}+\epsilon\frac{p_{\parallel}}{e}\nabla\times\unit{b}+\mathcal{O}(\epsilon^{2}).
\end{equation}
We will also find useful the expression
\begin{equation}
\frac{\mathbf{B}^{\star}}{B_{\parallel}^{\star}} \; \equiv \; \unit{b}+\epsilon\frac{p_{\parallel}}{eB_{\parallel}^{\star}}\unit{b}\times\bm{\kappa}+\mathcal{O}(\epsilon^2).
\end{equation}

The generating functions $G_{n}^{\alpha}$ that define the coordinate
transformations between between the guiding-center coordinates $Z^{\alpha}$
and particle coordinates $z^{\alpha}$ according to 
\begin{align}
Z^{\alpha}\;=\; z^{\alpha}+\epsilon\, G_{1}^{\alpha}+\epsilon^{2}\left(G_{2}^{\alpha}+\frac{1}{2}G_{1}^{\beta}\frac{\partial G_{1}^{\alpha}}{\partial z^{\beta}}\right)+\mathcal{O}(\epsilon^{2}),\\
z^{\alpha}\;=\; Z^{\alpha}-\epsilon\, G_{1}^{\alpha}-\epsilon^{2}\left(G_{2}^{\alpha}-\frac{1}{2}G_{1}^{\beta}\frac{\partial G_{1}^{\alpha}}{\partial Z^{\beta}}\right)+\mathcal{O}(\epsilon^{2}),
\end{align}
have the first order components for the spatial position and parallel
momentum 
\begin{align}
G_{1}^{\mathbf{X}}\;= & \;-\bm{\rho}_{0}\;\equiv\;-\sqrt{\frac{2m\mu}{e^{2}B}}\unit{\rho},\\
G_{1}^{p_{\parallel}}\;= & \;-p_{\parallel}\,\bm{\rho}_{0}\cdot\bm{\kappa}+\frac{m\mu}{e}\left(\tau_{B}+\mathsf{a}_{1}:\nabla\unit{b}\right),
\end{align}
as well as the components for the magnetic moment and gyroangle 
\begin{align}
G_{1}^{\mu}\;= & \;\bm{\rho}_{0}\cdot\left(\mu\nabla\ln B+\frac{p_{\parallel}^{2}}{mB}\bm{\kappa}\right)-\frac{\mu\, p_{\parallel}}{eB}\left(\tau_{B}+\mathsf{a}_{1}:\nabla\unit{b}\right),\\
G_{1}^{\theta}\;= & \;-\bm{\rho}_{0}\cdot\mathbf{R}+\frac{p_{\parallel}}{eB}\left(\mathsf{a}_{2}:\nabla\unit{b}\right)+\frac{\partial\bm{\rho}_{0}}{\partial\theta}\cdot\left(\nabla\ln B+\frac{p_{\parallel}^{2}}{2m\,\mu B}\bm{\kappa}\right),
\end{align}
where the magnetic field curvature vector is $\bm{\kappa}=\unit{b}\cdot\nabla\unit{b}$.
We also need the spatial component of the second order generating
function 
\begin{align}
G_{2}^{\mathbf{X}}\;= & \;\left[\frac{2p_{\parallel}}{eB}\left(\frac{\partial\bm{\rho}_{0}}{\partial\theta}\cdot\bm{\kappa}\right)+\frac{m\mu}{e^{2}B}\left(\mathsf{a}_{2}:\nabla\unit{b}\right)\right]\unit{b}
+\frac{p_{\parallel}}{eB}\,\tau_{B}\,\bm{\rho}_{0}\nonumber \\ & \;+\frac{1}{2}\left(G_{1}^{\mu}-\mu\bm{\rho}_{0}\cdot\nabla\ln B\right)\frac{\partial\bm{\rho}_{0}}{\partial\mu}+\frac{1}{2}\left(G_{1}^{\theta}+\bm{\rho}_{0}\cdot\mathbf{R}\right)\frac{\partial\bm{\rho}_{0}}{\partial\theta}.
\end{align}
The dyads $\mathsf{a}_{1}$ and $\mathsf{a}_{2}$ are 
\begin{align}
 & \mathsf{a}_{1}\;\equiv\;-\frac{1}{2}\left(\unit{\rho}\unit{\perp}+\unit{\perp}\unit{\rho}\right)\;=\;\frac{\partial\mathsf{a}_{2}}{\partial\theta},\\
 & \mathsf{a}_{2}\;\equiv\;\frac{1}{4}\left(\unit{\perp}\unit{\perp}-\unit{\rho}\unit{\rho}\right)\;=\;-\frac{1}{4}\frac{\partial\mathsf{a}_{1}}{\partial\theta}.
\end{align}

With the guiding-center Poisson bracket and Hamiltonian given, obtaining
the Hamiltonian equations of motion for each phase-space coordinate
is then straightforward. For the phase-space $Z^{\alpha}=(\mathbf{X},p_{\parallel},\mu,\theta)$
we find 
\begin{align}
\dot{\mathbf{X}}\;= & \;\{\mathbf{X},H_{gc}\}\;=\;\frac{p_{\parallel}}{\gamma m}\frac{\mathbf{B}^{\star}}{B_{\parallel}^{\star}}+\epsilon\frac{\unit{b}}{eB_{\parallel}^{\star}}\times\frac{\mu}{\gamma}\nabla B\\
\dot{p}_{\parallel}\;= & \;\{p_{\parallel},H_{gc}\}\;=\;-\frac{\mathbf{B}^{\star}}{B_{\parallel}^{\star}}\cdot\frac{\mu}{\gamma}\nabla B,\\
\dot{\mu}\;= & \;\{\mu,H_{gc}\}\;=\epsilon^{-1}\frac{e}{m}\frac{\partial H_{gc}}{\partial\theta}\;\equiv\;0,\\
\dot{\theta}\;= & \;\epsilon^{-1}\frac{eB}{\gamma m}+\dot{\mathbf{X}}\cdot\mathbf{R}^{\star}\;\equiv\;\epsilon^{-1}\Omega+\dot{\mathbf{X}}\cdot\mathbf{R}^{\star}.
\end{align}
While calculating the gyro-averages, one needs the expression 
\begin{align}
\nabla^{\star}\bm{\rho}_{0}\;= & \;-\frac{1}{2}\nabla\ln B\,\bm{\rho}_{0}+\frac{1}{2}\tau_{B}\unit{b}\frac{\partial\bm{\rho}_{0}}{\partial\theta}-\left(\nabla\unit{b}\cdot\bm{\rho}_{0}\right)\unit{b}
\end{align}
where $\nabla\times\unit{b}=\tau_{B}\unit{b}+\unit{b}\times\bm{\kappa}$,
and also 
\begin{equation}
\unit{b}\cdot\nabla^{\star}\bm{\rho}_{0}\;=\;\frac{\tau_{B}}{2}\frac{\partial\bm{\rho}_{0}}{\partial\theta}-\mu\,\nabla_{\parallel}\ln B\frac{\partial\bm{\rho}_{0}}{\partial\mu}-\left(\bm{\rho}_{0}\cdot\bm{\kappa}\right)\unit{b}.
\end{equation}

\section{Push-forward of the radiation reaction force}
\label{sec:push_forwards}
Noting that $\mathcal{T}_{gc}^{-1}\gamma\equiv\gamma$, the push-forwards
of particle momentum and magnetic field time derivative become 
\begin{align}
\mathcal{T}_{gc}^{-1}\mathbf{p}\;\equiv & \;\mathcal{T}_{gc}^{-1}\left(\gamma m\mathbf{v}\right)\;=\;\gamma m\left(\mathcal{T}_{gc}^{-1}\frac{d}{dt}\mathcal{T}_{gc}\right)\left(\mathcal{T}_{gc}^{-1}\mathbf{x}\right),\\
\mathcal{T}_{gc}^{-1}\dot{\mathbf{B}}\;\equiv & \;\left(\mathcal{T}_{gc}^{-1}\frac{d}{dt}\mathcal{T}_{gc}\right)\left(\mathcal{T}_{gc}^{-1}\mathbf{B}\right).
\end{align}
The guiding-center time derivative operator is 
\begin{align}
\left(\mathcal{T}_{gc}^{-1}\frac{d}{dt}\mathcal{T}_{gc}\right) \;\equiv\; \frac{\partial}{\partial t}+\dot{\mathbf{X}}\cdot\nabla+\dot{p}_{\parallel}\frac{\partial}{\partial p_{\parallel}}+\dot{\theta}\frac{\partial}{\partial\theta}
\;\equiv\; \frac{\partial}{\partial t}+\dot{\mathbf{X}}\cdot\nabla^{\star}+\dot{p}_{\parallel}\frac{\partial}{\partial p_{\parallel}}+\epsilon^{-1}\Omega\frac{\partial}{\partial\theta},
\end{align}
and because it involves a term of order $\epsilon^{-1}$, the push-forward of particle position 
\begin{align}
\mathcal{T}_{gc}^{-1}\mathbf{x} \;\equiv\; \mathbf{X}-\epsilon G_{1}^{\mathbf{X}}-\epsilon^{2}\left(G_{2}^{\mathbf{X}}-\frac{1}{2}G_{1}^{\beta}\frac{\partial G_{1}^{\mathbf{X}}}{\partial Z^{\beta}}\right)+\mathcal{O}(\epsilon^{3})
\;\equiv\; \mathbf{X}+\epsilon\bm{\rho}_{0}+\epsilon^{2}\bm{\rho}_{1}+\mathcal{O}(\epsilon^{3}),
\end{align}
and the push-forward of the magnetic field 
\begin{align}
\mathcal{T}_{gc}^{-1}\mathbf{B} \;\equiv\; \mathbf{B}-\epsilon G_{1}^{\mathbf{X}}\cdot\nabla\mathbf{B}-\epsilon^{2}\left[G_{2}^{\mathbf{X}}\cdot\nabla\mathbf{B}-\frac{1}{2}G_{1}^{\beta}\frac{\partial}{\partial Z^{\beta}}\left(G_{1}^{\mathbf{X}}\cdot\nabla\mathbf{B}\right)\right]+\mathcal{O}(\epsilon^{3}),
\end{align}
have to be evaluated up to second order in $\epsilon^{2}$. The explicit expression for the first order Larmor radius vector is 
\begin{align}
\bm{\rho}_{1}\;= & \;-\left[2\frac{p_{\parallel}}{eB}\left(\bm{\kappa}\cdot\frac{\partial\bm{\rho}_{0}}{\partial\theta}\right)-\frac{m\mu}{e^{2}B}\left(\mathsf{a}_{2}:\nabla\unit{b}-\frac{1}{2}\nabla\cdot\unit{b}\right)\right]\unit{b}\nonumber \\
 & \;-\left[\frac{1}{2}\frac{p_{\parallel}}{eB}\left(\tau_{B}-\mathsf{a}_{1}:\nabla\unit{b}\right)+\frac{\partial\bm{\rho}_{0}}{\partial\mu}\cdot\left(\mu\nabla\ln B+\frac{p_{\parallel}^{2}}{mB}\bm{\kappa}\right)\right]\bm{\rho}_{0}\nonumber \\
 & \;-\left[\frac{p_{\parallel}}{eB}\left(\mathsf{a}_{2}:\nabla\unit{b}\right)+\frac{\partial\bm{\rho}_{0}}{\partial\theta}\cdot\left(\nabla\ln B+\frac{p_{\parallel}^{2}}{2m\mu B}\bm{\kappa}\right)\right]\frac{\partial\bm{\rho}_{0}}{\partial\theta}.
\end{align}

Now, up to first order in $\epsilon$, we find the push-forward of
the particle momentum 
\begin{align}
\mathcal{T}_{gc}^{-1}\mathbf{p} \;= \;\gamma m\dot{\mathbf{X}}+eB\frac{\partial\bm{\rho}_{0}}{\partial\theta}+\epsilon\left(p_{\parallel}\unit{b}\cdot\nabla^{\star}\bm{\rho}_{0}+eB\frac{\partial\bm{\rho}_{1}}{\partial\theta}\right)+\mathcal{O}(\epsilon^{2})
\;\equiv\; \mathbf{p}_{0}+\epsilon\,\mathbf{p}_{1}+\mathcal{O}(\epsilon^{2}),
\end{align}
where the zeroth and first components are given by 
\begin{align}
\mathbf{p}_{0}\;= & \; p_{\parallel}\unit{b}+eB\frac{\partial\bm{\rho}_{0}}{\partial\theta},\\
\mathbf{p}_{1}\;= & \;\frac{\unit{b}}{eB_{\parallel}^{\star}}\times\left(m\mu\nabla B+p_{\parallel}^{2}\bm{\kappa}\right)+p_{\parallel}\unit{b}\cdot\nabla^{\star}\bm{\rho}_{0}+eB\frac{\partial\bm{\rho}_{1}}{\partial\theta}.
\end{align}
One also needs the push-forward of the radiation reaction time-scale
\begin{align}
\mathcal{T}_{gc}^{-1}\fr^{-1}\;= & \;\fr^{-1}\left(\frac{}{}1+2\,\epsilon\,\bm{\rho}_{0}\cdot\nabla\ln B\right)+\mathcal{O}(\epsilon^{2}),
\end{align}
the push-forward of the perpendicular momentum 
\begin{align}
\mathcal{T}_{gc}^{-1}\mathbf{p}_{\perp}\;= & \; eB\frac{\partial\bm{\rho}_{0}}{\partial\theta}+\epsilon\left(\mathbf{p}_{1}+G_{1}^{p_{\parallel}}\unit{b}-p_{\parallel}\bm{\rho}_{0}\cdot\nabla\unit{b}\right)+\mathcal{O}(\epsilon^{2})\\
\mathcal{T}_{gc}^{-1}p_{\perp}^{2}\;= & \;2m\mu B\left(1+\epsilon\bm{\rho}_{0}\cdot\nabla\ln B-\epsilon\frac{G_{1}^{\mu}}{\mu}\right)+\mathcal{O}(\epsilon^{2}),
\end{align}
and the push-forward of the magnetic field time-derivative 
\begin{align}
\mathcal{T}_{gc}^{-1}\dot{\mathbf{B}}\;= & \;\frac{p_{\parallel}}{\gamma m}\unit{b}\cdot\nabla\mathbf{B}+\Omega\frac{\partial\bm{\rho_{0}}}{\partial\theta}\cdot\nabla\mathbf{B}+\epsilon\frac{p_{\parallel}}{\gamma m}\unit{b}\cdot\nabla^{\star}\left(\bm{\rho}_{0}\cdot\nabla\mathbf{B}\right)\nonumber \\
 & \;-\epsilon\Omega\left[\frac{\partial G_{2}^{\mathbf{X}}}{\partial\theta}\cdot\nabla\mathbf{B}+\frac{1}{2}\frac{\partial}{\partial\theta}\left(G_{1}^{\beta}\frac{\partial}{\partial Z^{\beta}}\left(\bm{\rho}_{0}\cdot\nabla\mathbf{B}\right)\right)\right]\nonumber \\
 & \;+\epsilon\left[\frac{\unit{b}}{eB_{\parallel}^{\star}}\times\left(\frac{\mu}{\gamma}\nabla B+\frac{p_{\parallel}^{2}}{\gamma m}\bm{\kappa}\right)\right]\cdot\nabla\mathbf{B}+\mathcal{O}(\epsilon^{2}).
\end{align}
We also note that $\mathcal{T}_{gc}^{-1}\left(\fr^{-1}\Omega^{-1}B^{-1}\right)\equiv\fr^{-1}\Omega^{-1}B^{-1}$.

Combining the above expressions, we calculate the push-forward of the particle radiation reaction force 
\begin{align}
\mathcal{T}_{gc}^{-1}\mathbf{K}\;=&\;-\left(\mathcal{T}_{gc}^{-1}\fr^{-1}\right)\left[\left(\mathcal{T}_{gc}^{-1}\mathbf{p}_{\perp}\right)+\frac{\left(\mathcal{T}_{gc}^{-1}p_{\perp}^{2}\right)}{(mc)^{2}}\left(\mathcal{T}_{gc}^{-1}\mathbf{p}\right)\right]\nonumber \\
  &\;-\epsilon\fr^{-1}\Omega^{-1}B^{-1}\left(\mathcal{T}_{gc}^{-1}\dot{\mathbf{B}}\right)\times\left(\mathcal{T}_{gc}^{-1}\mathbf{p}\right).
\end{align}
Expressed as $\mathcal{T}_{gc}^{-1}\mathbf{K}\;=\mathbf{K}_{0}+\epsilon\mathbf{K}_{1}+\mathcal{O}(\epsilon^{2})$,
the zeroth order term is given by 
\begin{align}
\mathbf{K}_{0}\;= & \;-\fr^{-1}\left(eB\frac{\partial\bm{\rho}_{0}}{\partial\theta}+\frac{2\mu B}{mc^{2}}\mathbf{p}_{0}\right),
\end{align}
and for the first order term we have
\begin{align}
\mathbf{K}_{1}\;= & \;2\left(\bm{\rho}_{0}\cdot\nabla\ln B\right)\mathbf{K}_0-\fr^{-1}\Omega^{-1}B^{-1}\left(\frac{\mathbf{p}_{0}}{\gamma m}\cdot\nabla\mathbf{B}\right)\times\mathbf{p}_{0}\nonumber\\ & \; -\fr^{-1}\left[\left(1+\frac{2\mu B}{mc^{2}}\right)\mathbf{p}_{1}+G_{1}^{p_{\parallel}}\unit{b}-p_{\parallel}\bm{\rho}_{0}\cdot\nabla\unit{b} +\frac{2\mu B}{mc^{2}}\left(\bm{\rho}_{0}\cdot\nabla\ln B-\frac{G_{1}^{\mu}}{\mu}\right)\mathbf{p}_{0}\right].
\end{align}

\section{Gyro-averages of the guiding-center radiation reaction force}
\label{sec:gc_force}
With the push-forward of the particle phase-space \eero{RR-force} given in Appendix~\ref{sec:push_forwards}, we can evaluate the necessary gyro-averages with the help of a useful identity
\begin{align}
\nabla\unit{b}:\left\langle \unit{\perp}\unit{\rho}\right\rangle = & -\nabla\unit{b}:\left\langle \unit{\rho}\unit{\perp}\right\rangle =\frac{\tau_{B}}{2}.
\end{align}
This helps us evaluate the gyro-averages of the push-forwarded \eero{RR-force}:
\begin{align}
&\left\langle\mathbf{K}_0\right\rangle \; = \; -\fr^{-1}p_{\parallel}\,\frac{2\mu B}{mc^2}\,\unit{b},\\
&\left\langle\mathbf{K}_1\right\rangle \; = \; \fr^{-1}\frac{2\mu B}{mc^2}\frac{\unit{b}}{eB_{\parallel}^{\star}}\times\left(m\mu\nabla B+p_{\parallel}^2\bm{\kappa}\right) -3\fr^{-1}\frac{2\mu B}{mc^2}p_{\parallel}\unit{b}\times(\varrho_{\parallel}\bm{\kappa})-\fr^{-1}\frac{2\mu B}{mc^2}p_{\parallel}\varrho_{\parallel}\tau_B\unit{b},
\end{align}
as well as the rest of the gyro-averages that are needed in the expressions for $\mathcal{K}^{\mathbf{X}}$, $\mathcal{K}^{p_{\parallel}}$, and $\mathcal{K}^{\mu}$
\begin{align}
&\left\langle\nabla^{\star}\bm{\rho}_{0}\cdot\mathbf{K}_{0}\right\rangle \; = \; -\fr^{-1}\tau_B\frac{m\mu}{e}\left(1+\frac{2\mu B}{mc^2}\right)\unit{b},\\
&\left\langle\frac{\partial\bm{\rho}_{1}}{\partial p_{\parallel}}\cdot\mathbf{K}_{0}\right\rangle \; = \; 0,\\
&\left\langle\frac{\partial\bm{\rho}_0}{\partial\theta}\cdot\mathbf{K}_0\right\rangle\; = \; - 2\fr^{-1}\frac{m\mu}{e}\left(1+\frac{2\mu B}{mc^2}\right),\\
&\left\langle \frac{\partial\bm{\rho}_{0}}{\partial\theta}\cdot\mathbf{K}_{1}\right\rangle \;= \; -2\fr^{-1}\frac{m\mu}{e}\left(1+\frac{2\mu B}{mc^2}\right)\,\varrho_{\parallel}\tau_B,\\
&\left\langle \frac{\partial\bm{\rho}_{1}}{\partial\theta}\cdot\mathbf{K}_{0}\right\rangle \; = \; \;\fr^{-1}\frac{m\mu}{e}\left(1+\frac{2\mu B}{mc^{2}}\right)\varrho_{\parallel}\tau_{B}.
\end{align}

Now, the spatial component of the guiding-center radiation reaction force becomes
\begin{align}
\mathcal{K}^{\mathbf{X}}\;= & \;-\epsilon\frac{\unit{b}}{eB_{\parallel}^{\star}}\times\left\langle\mathbf{K}_{0}+\epsilon\mathbf{K}_{1}+\epsilon\nabla^{\star}\bm{\rho}_{0}\cdot\mathbf{K}_{0}\right\rangle-\epsilon^2\unit{b}\left\langle\frac{\partial\bm{\rho}_{1}}{\partial p_{\parallel}}\cdot\mathbf{K}_{0}\right\rangle,\nonumber\\
= &\; -\epsilon^2\,\frac{\fr^{-1}}{\Omega_{\parallel}^{\star}}\frac{2\mu B}{mc^2}\left(\unit{b}\times\dot{\mathbf{X}}+3v_{\parallel}\,\varrho_{\parallel}\,\bm{\kappa}\right),
\end{align}
where we have introduced the modified gyro frequency 
\begin{align} 
\Omega_{\parallel}^{\star}=(eB_{\parallel}^{\star})/(\gamma m)=\Omega\,(1+\epsilon\varrho_{\parallel}\tau_B).
\end{align}
 For the parallel momentum component we find
\begin{align}
\mathcal{K}^{p_{\parallel}}\;= & \;\frac{\mathbf{B}^{\star}}{B_{\parallel}^{\star}}\cdot\Big\langle\mathbf{K}_{0}\Bigr\rangle+\epsilon\unit{b}\cdot\left\langle\mathbf{K}_1+\nabla^{\star}\bm{\rho}_{0}\cdot\mathbf{K}_{0}\right\rangle,\nonumber\\
= & \; -\fr^{-1}p_{\parallel}\frac{\mu B}{mc^2}\left(2+\epsilon\varrho_{\parallel}\tau_B\right)-\epsilon\fr^{-1}\frac{p_{\perp}\gamma^2}{2}\varrho_{\perp}\tau_B,
\end{align}
and for the magnetic moment $\mu$ the force becomes 
\begin{align}
\mathcal{K}^{\mu}\;= & \;\frac{e}{m}\left\langle \frac{\partial\bm{\rho}_{0}}{\partial\theta}\cdot\Bigl(\mathbf{K}_{0} +\epsilon\mathbf{K}_{1}\Bigr)\right\rangle +\epsilon\frac{e}{m}\left\langle \frac{\partial\bm{\rho}_{1}}{\partial\theta}\cdot\mathbf{K}_{0}\right\rangle,\nonumber\\
= & \; -\fr^{-1}\mu\left(1+\frac{2\mu B}{mc^2}\right)\left(2+\epsilon\varrho_{\parallel}\tau_B\right).
\end{align}

\bibliographystyle{unsrt}
\bibliography{revision}

\providecommand{\noopsort}[1]{}\providecommand{\singleletter}[1]{#1}
\begin{thebibliography}{10}

\bibitem{lorentz}
H.A. Lorentz.
\newblock La {T}h\'{e}orie \'{E}lectromagn\'{e}tique de {Maxwell} et {S}on
  {Application} {Aux} {Corps} {Mouvants}.
\newblock In {\em Collected Papers}, pages 164--343. Springer Netherlands,
  1936.

\bibitem{abraham}
M.~Abraham.
\newblock {\em Theorie der Elektrizit\"{a}t, Vol II: Elektromagnetische Theorie
  der Strahlung}.
\newblock Teubner Leipzig, 1905.

\bibitem{dirac:1938}
P.~A.~M. Dirac.
\newblock Classical theory of radiating electrons.
\newblock {\em Proceedings of the Royal Society of London. Series A,
  Mathematical and Physical Sciences}, 167(929):pp. 148--169, 1938.

\bibitem{pauli1958theory}
W.~Pauli.
\newblock {\em Theory of Relativity}.
\newblock Dover Books on Physics. Dover Publications, 1958.

\bibitem{landau_lifshitz_fields}
L.~D. Landau and E.~M. Lifshitz.
\newblock {\em The Classical Theory of Fields}, volume~2 of {\em Course of
  Theoretical Physics}.
\newblock Pergamon, Amsterdam, fourth edition edition, 1975.

\bibitem{Bakhtiari_et_al:PRL2005}
M.~Bakhtiari, G.~J. Kramer, M.~Takechi, H.~Tamai, Y.~Miura, Y.~Kusama, and
  Y.~Kamada.
\newblock Role of bremsstrahlung radiation in limiting the energy of runaway
  electrons in tokamaks.
\newblock {\em Physical Review Letters}, 94:215003, Jun 2005.

\bibitem{adam:effective_field_PRL}
A.~Stahl, E.~Hirvijoki, J.~Decker, O.~Embr\'eus, and T.~F\"ul\"op.
\newblock Effective critical electric field for runaway-electron generation.
\newblock {\em Physical Review Letters}, 114:115002, Mar 2015.

\bibitem{littlejohn:jpp:4732464}
Robert~G. Littlejohn.
\newblock Variational principles of guiding centre motion.
\newblock {\em Journal of Plasma Physics}, 29:111--125, 2 1983.

\bibitem{RevModPhys.81.693}
John~R. Cary and Alain~J. Brizard.
\newblock Hamiltonian theory of guiding-center motion.
\newblock {\em Reviews of Modern Physics}, 81:693--738, May 2009.

\bibitem{andersson:pop2001}
F.~Andersson, P.~Helander, and L.-G. Eriksson.
\newblock Damping of relativistic electron beams by synchrotron radiation.
\newblock {\em Physics of Plasmas}, 8(12):5221--5229, 2001.

\bibitem{guan:2010}
Xiaoyin Guan, Hong Qin, and Nathaniel~J. Fisch.
\newblock Phase-space dynamics of runaway electrons in tokamaks.
\newblock {\em Physics of Plasmas}, 17(9):092502, 2010.

\bibitem{liu:pop2014}
Jian Liu, Hong Qin, Nathaniel~J. Fisch, Qian Teng, and Xiaogang Wang.
\newblock What is the fate of runaway positrons in tokamaks?
\newblock {\em Physics of Plasmas}, 21(6):--, 2014.

\bibitem{rohrlich2007classical}
F.~Rohrlich.
\newblock {\em Classical Charged Particles}.
\newblock World Scientific, 2007.

\bibitem{griffiths:AJP:2009}
David~J. Griffiths, Thomas~C. Proctor, and Darrell~F. Schroeter.
\newblock Abraham--lorentz versus landau--lifshitz.
\newblock {\em American Journal of Physics}, 78(4):391--402, 2010.

\bibitem{Ford1993182}
G.W Ford and R.F O'Connell.
\newblock Relativistic form of radiation reaction.
\newblock {\em Physics Letters A}, 174(3):182 -- 184, 1993.

\bibitem{Spohn:EPL:2000}
H.~Spohn.
\newblock The critical manifold of the lorentz-dirac equation.
\newblock {\em Europhysics Letters}, 50(3):287, 2000.

\bibitem{brizard:4429}
A.~J. Brizard.
\newblock A guiding-center {Fokker}--{Planck} collision operator for nonuniform
  magnetic fields.
\newblock {\em Physics of Plasmas}, 11(9):4429--4438, 2004.

\bibitem{brizard:2009}
A.~J. Brizard, J.~Decker, Y.~Peysson, and F.-X. Duthoit.
\newblock {Orbit}-averaged guiding-center {Fokker}--{Planck} operator.
\newblock {\em Physics of Plasmas}, 16(10), 2009.

\bibitem{decker:2010}
J.~Decker, Y.~Peysson, A.~J. Brizard, and F.-X. Duthoit.
\newblock {Orbit}-averaged guiding-center {Fokker}--{Planck} operator for
  numerical applications.
\newblock {\em Physics of Plasmas (1994-present)}, 17(11), 2010.

\end{thebibliography}

\end{document}